\documentclass[pre,aps,amsmath,amssymb,twocolumn,superscriptaddress]{revtex4-1} 
\usepackage[english]{babel}
\usepackage[colorlinks=true,citecolor=blue]{hyperref}
\usepackage{amsmath,amssymb}
\usepackage[usenames]{color}
\usepackage{grffile}
\usepackage[pdftex]{graphicx}
\usepackage{amsmath, amstext, amssymb, amsfonts, amsxtra}
\usepackage{textcomp}
\usepackage{xspace}
\DeclareSymbolFontAlphabet{\amsmathbb}{AMSb} 
\usepackage{soul}
\usepackage{xcolor}
\usepackage{epstopdf}
\usepackage{braket}
\usepackage{float}
\usepackage{tikz}
\usepackage{enumitem}
\setlist{nosep}

\newcommand{\bop}{\hat{b}}

\newcommand{\rhop}{\hat{\rho}} 
\newcommand{\sop}{\hat{\sigma}}

\newcommand{\tr}{{\rm tr}}

\newcommand{\Hop}{\hat{H}}

\newcommand{\Sop}{\hat{S}}     
\newcommand{\jop}{\hat{j}}    
\newcommand{\hop}{\hat{h}}    
\newcommand{\Rop}{\mathcal{R}}

\newcommand{\smb}[1]{\left(#1\right)}
\newcommand{\sbr}[1]{\left[#1\right]}

\newcommand{\cmute}[2]{\left[{#1},{#2}\right]}
\newcommand{\ii}{\mathrm{i}}
\newcommand{\expe}{\mathrm{e}}

\newcommand{\sutdepd}{EPD Pillar, Singapore University of Technology and Design, 8 Somapah Road, 487372 Singapore} 
\newcommand{\sutdsci}{Science and Math Cluster, Singapore University of Technology and Design, 8 Somapah Road, 487372 Singapore} 
\newcommand{\zth}{Department of Physics, University of Zurich, Winterthurerstrasse 190, 8057 Zurich, Switzerland}
\newcommand{\majulab}{MajuLab, CNRS-UCA-SU-NUS-NTU International Joint Research Unit, 117543 Singapore}

\begin{document}

\title{Transport and energetic properties of a ring of interacting spins coupled to heat baths} 

\author{Xiansong Xu} 
\affiliation{\sutdsci}
\author{Kenny Choo}
\affiliation{\zth}
\author{Vinitha Balachandran} 
\affiliation{\sutdepd}
\author{Dario Poletti}
\affiliation{\sutdsci} 
\affiliation{\sutdepd} 
\affiliation{\majulab}

\begin{abstract} 
We study the heat and spin transport properties in a ring of interacting spins coupled to heat baths at different temperatures. We show that interactions, by inducing avoided crossings, can be a means to tune both the total heat current flowing between the ring and the baths, and the way it flows through the system. In particular, we recognize three regimes in which the heat current flows clockwise, counterclockwise, and in parallel. The temperature bias between the baths also induces a spin current within the ring, whose direction and magnitude can be tuned by the interaction. Lastly, we show how the ergotropy of the nonequilibrium steady state can increase significantly near the avoided crossings. 
\end{abstract}

\maketitle
\section{\label{Sec:intro} Introduction}

Understanding transport properties in quantum systems can lead to various interesting applications such as quantum rectifiers \cite{Segal2005,Chang2006,Kobayashi2009,Arrachea2009,Wu2009,Zhang2009,Yan2009,Werlang2014,Martinez-Perez2015, Balachandran2018, BalachandranPoletti2018, BalachandranPoletti2018a, Motz2018}, transistors \cite{Joulain2016}, engines \cite{Kosloff2014,Uzdin2015, Gelbwaser-Klimovsky2015, AltintasMustecaplioglu2015, BenentiWhitney2017, Bissbort2017, Seah2018a, Roulet2018, HovhannisyanImparato2018, Blickle2012,Martinez2016,Serra-Garcia2016, Rossnagel2016,KlatzowPoem2017,VanHorneMukherjee2018,vonLindenfelsPoschinger2018, PetersonSerra2018}, refrigerators \cite{Levy2012, Linden2010, Mitchison2016, MaslennikovMatsukevich2017, Mu2017, Seah2018}, and batteries \cite{Campaioli2017, Ferraro2018, Andolina2018}. 
In thermoelectric systems, one main goal is to achieve an electric current due to a temperature bias. It is thus important to study fundamental aspects of the conversion of heat currents into particle/spin currents. In the following, we will consider a circuit made of a ring of spins which we couple to baths at different temperatures to study whether a spin current can be induced by them and how this current can qualitatively change depending on the system parameters. We will evaluate the maximum amount of work that the steady state can produce by computing its ergotropy \cite{Allahverdyan2004}.      
In particular, we focus on the interplay between interactions in the system and the external driving due to the two different thermal baths. To find more generic results, we focus on a minimal model as the study of minimal models allows one to extract the key ingredients necessary to obtain a given physical phenomenon.

\begin{figure}[H]
\centering
\includegraphics[width=0.85\columnwidth]{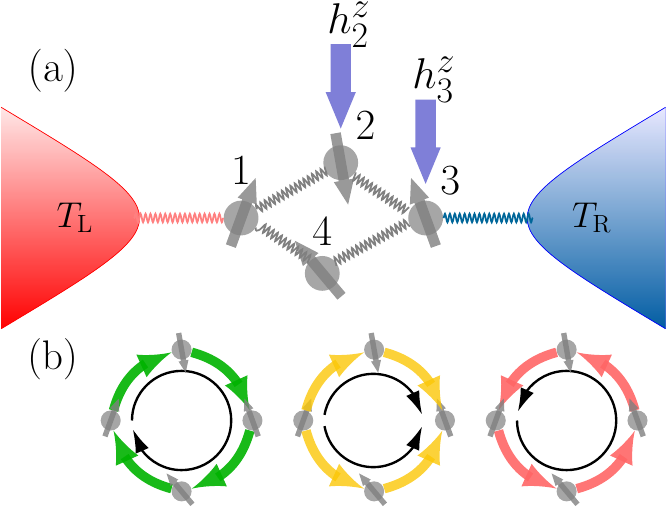}
\caption{(\textbf{a}) schematic representation of a ring of interacting spins with local magnetic fields coupled to two heat baths; (\textbf{b}) three possible heat current modes in the system: clockwise, parallel, and counterclockwise. Since there is no spin exchange with the baths, the spin current can only flow in clockwise or counterclockwise directions.}
\label{fig:ring}
\end{figure}

The minimal model we consider is a ring of four spin $1/2$ with XXZ interaction \footnote{this can be mapped to spinless fermions with nearest neighbor interaction via Jordan--Wigner transformation}. The ring is coupled at two opposite sites to two baths at different temperatures, as shown in Figure \ref{fig:ring}(a). In order for the current to have a preferred direction, we need to break the symmetries of the model, which can be done by applying different local magnetic fields. 
The ring only exchanges energy in the form of heat with the baths and there is no transfer of spins between them. In this way, we can clearly analyze whether temperature biases can generate a spin current.

There can be two types of currents in the system: the heat current and the spin current.  
As pictorially represented in Figure \ref{fig:ring}(b), we will show that heat currents can flow in three different ways: the heat can flow in a parallel manner (yellow arrows) in the upper and lower part of the ring, or it can flow in an anti-parallel fashion. In the latter case, the flow can be either clockwise (green arrows) or counterclockwise (red arrows). As for the spin current, since no spin is exchanged with the baths, it can only flow clockwise or counterclockwise. We will show that, in our setup, the spin current inversion occurs together with a significant change in the heat exchanged with the baths. They both occur in the proximity of an avoided crossing which is induced by the interaction between the spins. Since the steady state is not in equilibrium, we also investigate the maximum energy that can be extracted from it via unitary processes, i.e., the ergotropy \cite{Allahverdyan2004}. We show what is the main contributor to the ergotropy of the steady state and also find that it is significantly enhanced near interaction induced avoided crossings.

The article is organized as follows: in Section \ref{Sec:model}, we briefly describe the interacting spin ring model coupled to heat baths. In Section \ref{Sec:method}, we discuss the Markovian Redfield master equation which we use to derive our results. In Section \ref{Sec:mode}, we study how the interaction can result in different scenarios of local heat and spin currents and, in Section \ref{Sec:ergotropy}, we analyze its effects on the ergotropy. In Section \ref{Sec:conclusions}, we draw our conclusions.      

\section{Model}
\label{Sec:model}

{We aim to study the effect of interactions on both the heat and spin currents. We thus consider a prototypical model, i.e., an XXZ chain with periodic boundary conditions.}  
The Hamiltonian is given~by 
\begin{align}
	\Hop_\mathrm{S} = & \sum_{l=1}^{4}\left[J \left(\sop^{x}_l \sop^x_{l+1}+\sop^y_l \sop^y_{l+1}\right)+\Delta \sop^u_l \sop^u_{l+1}\right] \nonumber\\  
        & + h^z_2 \sop^z_2 + h^z_3 \sop^z_3 ,
  \label{eq: Ham}
\end{align}
where $\sop^{u}_l=(\sop^z_l+1)/2$ {and where site $l=5$ coincides with site one} \footnote{We have written the interaction as $\sop^u_l\sop^u_l$ so that, when converting the spins to fermionic particles, this interaction term corresponds to only nearest neighbor interaction without a $\Delta$-dependent local chemical potential}. The $\sop^a_l$ with $a=x,y,z$ are the operators corresponding to the Pauli matrices.   

In order to generate internal currents in the system, it is necessary to break the reflection and rotational symmetries. This can be done by applying an external perturbation, or disorder, in terms of local magnetic field. The minimum type of perturbation needed to break the symmetries is to apply a different local magnetic field to two consecutive spins. Here, local magnetic fields $h_l^z$ are applied to the second and third spins as depicted in Figure \ref{fig:ring}(a). 
We set $\hbar = k_{\rm B} = J = 1$ throughout.

The two spins at sites $l=1,\;3$ are coupled to two heat baths as shown in Figure \ref{fig:ring}(a). The left and right heat baths are infinite collections of harmonic oscillators with their respective Hamiltonians $\Hop_{\rm L}$ and $\Hop_{\rm R}$ given by
\begin{align} 
  \Hop_{\rm L/R} &= \sum^\infty_{\omega_{\rm L/R}=0} \!\!\omega_{\rm L/R} \;  \bop^\dagger_{\omega_{\rm L/R}} \bop_{\omega_{\rm L/R}}.
  \label{eq:Hb}
\end{align}

We consider baths characterized by the spectral density given by ${\rm J}(\omega) = \gamma\omega \exp \left( - \omega/ \omega_{\rm c} \right)$, where $\omega_{\rm c}$ is the cutoff frequency and $\gamma$ is the system-bath coupling strength. We use a cutoff of $\omega_{\rm c} =10$ for both baths. The system-bath coupling strength $\gamma$ is chosen as $0.01$, which is sufficiently weak as compared to the level spacings, even in the presence of avoided crossings.     

The ring and the baths interact via system operators $\sop^z_{1}$ and $\sop^z_{3}$  as
  \begin{align}\label{eq:sys_bath}
    \Hop_{\rm SB} = &\sum_{\omega_{\rm L}=0}^\infty \sqrt{{\rm J}(\omega_{\rm L})}\sop^z_{\rm 1}  \left( \bop^\dagger_{\omega_{\rm L}} + \bop_{\omega_{\rm L}} \right) \nonumber \\
    & + \sum_{\omega_{\rm R}=0}^\infty \sqrt{{\rm J}(\omega_{\rm R})}\sop^z_{\rm 3} \left( \bop^\dagger_{\omega_{\rm R}} + \bop_{\omega_{\rm R}} \right). 
\end{align}

Since the operator $\sop^z_l$ conserves the total number of spins in the system, there is no spin current between the baths and the system. However, as we show later, the heat current can induce a spin current within the system in a direction that depends on the strength of the interaction.

\subsection*{Minimality of the Model} \label{sec:minimal}
We have considered a circuit of four spins, and this is the minimal number required to observe the effects of interaction on the ring current. With two spins, it is not possible to form a close circuit. With three spins, it is possible to form the smallest circuit; however, in such a system, the interaction term would be equivalent to a global energy shift. This can be shown quite simply: since the total magnetization is conserved, for three spins, it is only possible to have a current only with one spin up and two spins down or vice versa. However, in either case, the number of possible configurations is three and in all configurations, the number of parallel and anti-parallel neighboring spins is invariant. Hence, the interaction is effectively a homogeneous local potential. With four spins instead, in the magnetization sector with two spins up and two spins down, it is possible to have configurations in which the number of parallel or anti-parallel neighboring spins is different, e.g., alternating spins up and down or two spins up followed by two spins down. Hence, four spins is the smallest size of a circuit which can be used to explore the effects of interactions.       

\section{Methods}
\label{Sec:method}
To study the nonequilibrium properties of the steady state, we use the Redfield master equation~\cite{Redfield1957} which can be derived from Eqs. (\ref{eq: Ham})--(\ref{eq:sys_bath}). The Redfield master equation is a second-order perturbative master equation that produces accurate results for the occupation of the energy levels at zeroth order in system-bath coupling. The off-diagonal elements in the energy eigenbasis are, on the other hand, accurate up to the second order \cite{Fleming2011, Thingna2012, Thingna2013}. One important advantage of using this master equation is that it does not require the secular approximation which can result in the vanishing local currents within the system \cite{Purkayastha2016, WichterichMichel2007, Xu2017a, Rivas2017}. The Redfield master equation is often criticized for producing negative probabilities \cite{Ishizaki2009a}. We emphasize here that a strong Markovian environment, i.e., a fast decaying correlation function, usually helps to avoid such a problem. Moreover, the emergence of negative probabilities is a clear warning that the master equation is being used beyond its regime of validity (a check which is not present in master equations in Gorini--Kossakowski--Sudarshan--Lindblad (GKSL) form \cite{GoriniSudarshan1976, Lindblad1976}). In the simulations presented here, we have not observed negative probabilities. 

To compute the nonequilibrium steady-state (NESS) density operator $\rhop_{\rm NESS}=\rhop(t=\infty)$, we use the Redfield master equation given by \cite{BreuerPetruccione2007, Redfield1957},  
\begin{align}\label{eq:RME}
  \frac{d\rhop(t)}{dt}=&-{\ii}\cmute{\Hop_{\rm s}}{\rhop\smb{t}}+\sum_l \Rop_l \left[\rhop(t)\right],  
\end{align}	
where the dissipative part $\Rop_{l}$ are  
\begin{align}\label{eq:RME_S} 
  \Rop_{l} \left[\rhop(t)\right]=& \cmute{\hat{\mathbb{S}}_l\rhop(t)}{\Sop_l}+\cmute{\Sop_l}{\rhop(t)\hat{\mathbb{S}}_l^\dagger }, 
\end{align}	
with
\begin{align}\label{eq:S_op} 
  \hat{\mathbb{S}}_l=&\int^\infty_0 d\tau e^{-\mathrm{i} \Hop_{\mathrm{S}}\tau} \Sop_l e^{\mathrm{i} \Hop_{\mathrm{S}}\tau} C_l(\tau),  
\end{align} 
and where, for our setup, $\Sop_{\rm L}=\sop^z_1$ and $\Sop_{\rm R}=\sop^z_3$, and we associate the subindex ${\rm L}$, of the left bath, with site $l=1$ and  subindex ${\rm R}$, of the right bath, with site $l=3$. 
Since the upper limit of the integral in Eq. (\ref{eq:S_op}) is $\tau=\infty$, Eq. (\ref{eq:RME}) is also referred to as the Markovian Redfield master equation. Note that the action of the baths on the system is considered to be additive, which is in general a good approximation for sufficiently weak coupling and Markovian baths \cite{Kolodynski2018}.    
It should also be remarked that this approach considers the full Hamiltonian of the system $\Hop_{\rm S}$ in deriving $\hat{\mathbb{S}}_l$, and not a local approximation of it. This is important because the local Hamiltonian approximation can result in a failure to capture the dependence on many-body interactions \cite{XuPoletti2019a} and can be thermodynamically inconsistent \cite{Levy2014}. 
In Eqs. (\ref{eq:RME_S}) and (\ref{eq:S_op}), $\Rop_l\left[\cdot\right]$ is the dissipator that contains all the bath information through the transition operator $\hat{\mathbb{S}}_l$. 
The bath correlator $C_l(\tau)$ is explicitly given by 
\begin{align}
  C_l(\tau)=\int^\infty_0 \frac{d\omega}{\pi} \mathrm{J}\smb{\omega}\sbr{\coth\smb{\frac{\omega}{2T_l}}{\cos\smb{\omega \tau}}-\ii\sin\smb{\omega\tau}},
\end{align}
where ${{\rm J}(\omega)}$ is the spectral density specified in Section \ref{Sec:model} \cite{BreuerPetruccione2007} and $T_l$ is the temperature in either the left ($T_{\rm L}$) or right ($T_{\rm R}$) bath (see also Figure \ref{fig:ring}). 

The Markovian Redfield master equation in the energy eigenbasis results in the following equation for the density matrix $\rho_{\alpha,\beta}$, 
\begin{align} \label{eq:mr_matrices}
	\frac{d\rho_{\alpha,\beta}}{dt}  = & -{\rm i} \Delta_{\alpha,\beta} \rho_{\alpha,\beta} + \sum_{l={\rm L,R}}\;\sum_{\alpha',\beta'} \mathcal{R}_{l,\alpha,\beta}^{\alpha',\beta'} \; \rho_{\alpha',\beta'}, 
\end{align} 
where $\Delta_{\alpha,\beta}=E_{\alpha}-E_{\beta}$ is the energy difference between the energy levels $\alpha$ and $\beta$, while $\mathcal{R}^{\alpha',\beta'}_{l,\alpha,\beta}$ is a tensor acting on the density matrix $\rho_{\alpha,\beta}$ given by  
\begin{align} 
  \mathcal{R}^{\alpha',\beta'}_{l,\alpha,\beta} = &  \sum_{\alpha',\beta'} {\Big (} \mathbb{S}_{l,\alpha,\alpha'} S_{l,\beta,\beta'} + S_{l,\alpha,\alpha'}\mathbb{S}^\dagger_{l,\beta,\beta'} \nonumber \\ 
   &-\delta_{\beta,\beta'}\sum_\nu S_{l,\alpha,\nu}\mathbb{S}_{l,\nu,\alpha'} \nonumber \\
   &-\delta_{\alpha,\alpha'}\sum_{\nu}\mathbb{S}^\dagger_{l,\beta',\nu}S_{l,\nu,\beta}\Big),
    \label{eq: eigenRME}
\end{align}
with $\delta_{\alpha,\beta}$ the Kronecker delta. Here, $\mathbb{S}^\dagger_{l,\beta,\beta'}$ and $S^\dagger_{l,\beta,\beta'}$ are derived, respectively, from the coefficients of the operators $\hat{\mathbb{S}}_l$ and $\Sop_l$ in the energy eigenbasis.     

From Eq. (\ref{eq:S_op}), we get a transition matrix in the energy basis given by 
\begin{align}\label{eq:S_lab}
  \mathbb{S}_{l,\alpha,\beta}=&\int^\infty_0 d\tau e^{-\mathrm{i} \Delta_{\alpha,\beta} \tau} S_{l,\alpha,\beta}  C_l(\tau),  
\end{align}
which can be evaluated using Plemelj formula $\int^\infty_0 dt \expe^{\pm\ii\epsilon t} = \pi \delta\smb{\epsilon} \pm \mathrm{i}\frac{P}{\epsilon}$. The imaginary part, usually known as the Lamb shift, is a small perturbation to the system Hamiltonian and it is therefore neglected.

In case the secular approximation had been taken, as for master equations in GKSL form, then all the terms in $\mathcal{R}_{l,\alpha,\beta}^{\alpha',\beta'}$ in Eq.~\eqref{eq: eigenRME}, with $\alpha\ne \alpha'$ and $\beta\ne \beta'$ will be omitted. As a consequence, the resultant master equation would give vanishing off-diagonal elements for the reduced density matrix \cite{BlumBlum2012a, Xu2017a}.

\section{Local current modes}
\label{Sec:mode}

We are interested in the local heat and spin currents within the system. 
To obtain the expressions for the spin and heat currents, we use the conservation laws of, respectively, the local spin $\sop^u_l$ and the local energy on a bond $\hop_{l,l+1}$ where 
\begin{align}\label{eq:ham_loc}
\hop_{l,l+1}=& J \left(\sop^x_l\sop^x_{l+1} + \sop^y_l\sop^y_{l+1}\right) + \Delta \sop^u_l\sop^u_{l+1} \nonumber \\ 
& + \frac {h_2^z} 2  (\delta_{l,2}+\delta_{l+1,2}) + \frac {h_3^z} 2   (\delta_{l,3}+\delta_{l+1,3}).  
\end{align}

Hence, the local heat current operator on the $l$-th site, except for the sites in contact with the baths, is given by
\begin{align}\label{eq:jH}
  \jop_l^{\rm \; H}=\ii\left[\hat{h}_{l-1,l}, ~\hat{h}_{l,l+1} \right].
\end{align}

Similarly, for the local spin current operator, we have
\begin{align}
  \hat{j}_l^{\rm \; S}=J\left(\sop^x_l \sop^y_{l+1} - \sop^y_l \sop^x_{l+1} \right), 
\end{align}
which is uniform through the system because the total magnetization is conserved, and it can only be in the clockwise or counterclockwise direction.

In our setup, the heat current in the upper part of the ring is given by $j_{\rm 2}^{\rm H}=\braket{\hat{j}_2^{\rm \; H}}$, while the one in the lower part of the ring is given by $j_{\rm 4}^{\rm H}=\braket{\hat{j}_4^{\rm \; H}}$. Here and henceforth, we use the notation $\braket{\cdot}=\tr\left(~\cdot~\rhop_{\rm NESS}\right)$ to indicate the trace over the steady-state density operator of the system $\rhop_{\rm NESS}$. 
The total heat current is given by the difference of the current in the upper and lower part of the ring \footnote{Current operators are defined as positive in the clockwise direction}, and it is also given by the energy exchanged with each bath, which is computed by $j^{\rm H}_{\rm L}=\tr\left(\Hop_{\rm S} \Rop_{\rm L}[\rhop_{\rm NESS}]\right)$, i.e., the energy drawn from the left bath, or equivalently from $j^{\rm H}_{\rm R}=\tr\left(\Hop_{\rm S} \Rop_{\rm R}[\rhop_{\rm NESS}]\right)$, i.e., the energy drawn from the right bath. Indeed, we have that $j^{\rm H}_{\rm L}=-j^{\rm H}_{\rm R}=j_{\rm 2}^{\rm H}-j_{\rm 4}^{\rm H}$. 
We use $j^{\rm \;S} = \braket{\hat{j}_l^{\rm \;S}}$ to denote the local spin current. 

Due to the conservation of total magnetization (number of spins up), the system has multiple invariant subspaces. When there is zero or one spin up, the interaction $\Delta$ does not play any role in the system. When there are three or four spin ups, the interaction $\Delta$ in Eq. (\ref{eq: Ham}) acts effectively as local fields. For two spins up, $\Delta$ differentiates cases in which the two spins are next to each other or not, hence acting as a nearest neighbor interaction. We thus study the system in the symmetry sector with two spins up and contrast it with results from the sector with three spins up.

\begin{figure}
	\centering 
        \includegraphics[width=\columnwidth]{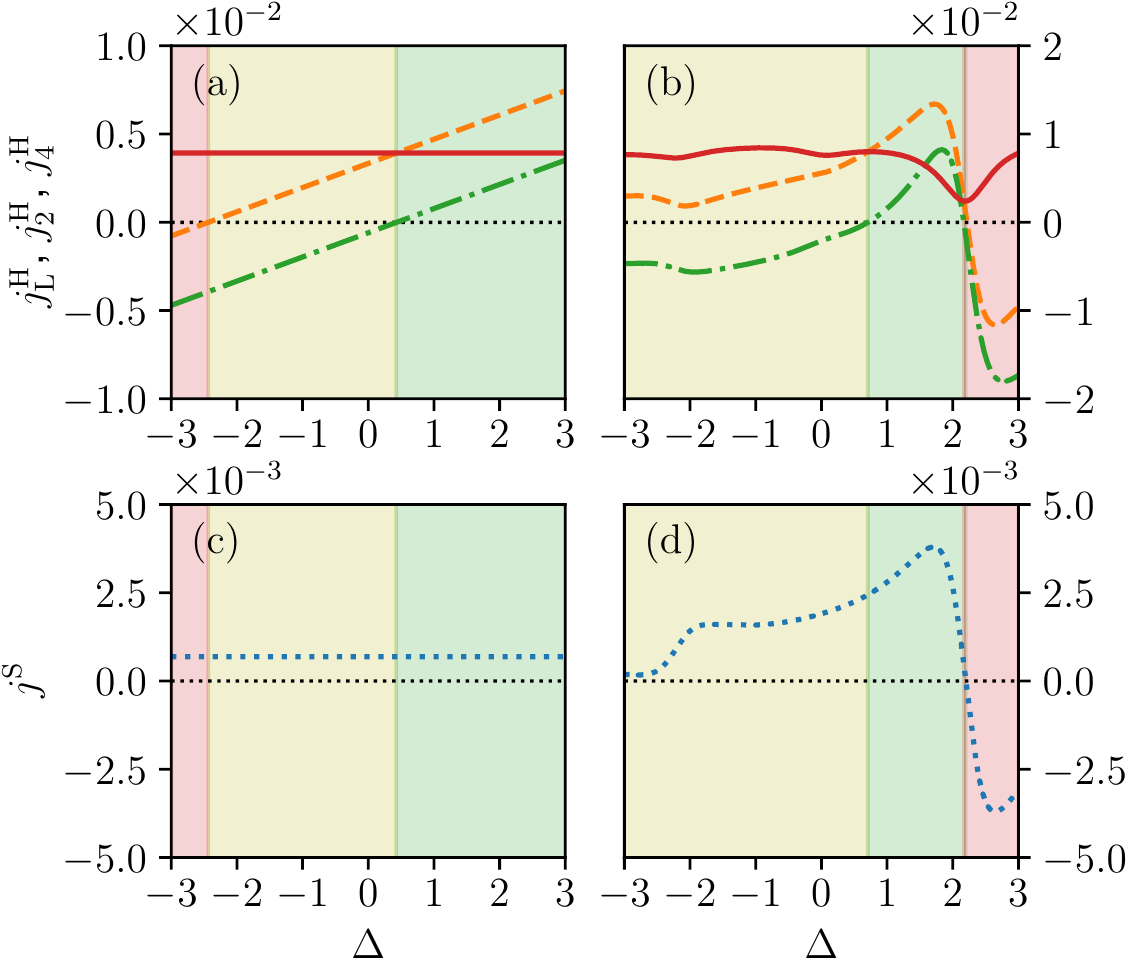}
        \caption{Changes of local energy current on the upper branch $j^{\rm H}_2$ (orange dashed line) and lower branch $j^{\rm H}_4$ (green dot-dashed line), and total energy current $j^{\rm H}_{\rm L}$ (red solid line) for (\textbf{a}) three spins up and for (\textbf{b}) two spins up; Spin current $j^{\rm \;S}$ (blue dotted line), as a function interaction strength $\Delta$ for (\textbf{c}) three spins up and for (\textbf{d}) two spins up. The shadings represent the regions of parameters for which heat current is counterclockwise (pink shading), parallel (yellow shading) and clockwise (green shading). The temperature of the left and right bath are $T_{\rm L}=2$ and $T_{\rm R}=1$, respectively. System-bath coupling strength $\gamma=0.01$ and local magnetic fields values are $h^z_2=2$ and $h^z_3=1$.} 
        \label{fig:Fig2}
\end{figure}

In Figure \ref{fig:Fig2}(a), we consider the case with three spins up. The local heat currents $j^{\rm H}_{2}$ and $j^{\rm H}_{4}$ are depicted by the orange dashed and green dot-dashed lines, respectively. While the total heat current (red solid line) remains unchanged when $\Delta$ increases, the local heat currents change linearly with $\Delta$ because of the linear shift of the eigenenergies. Because of this linear change, the local heat currents geometry changes from counterclockwise (pink shaded region in Figure \ref{fig:Fig2}) to parallel (yellow shaded region) and finally to clockwise (green shaded region). The local spin currents, on the other hand, do not vary with respect to $\Delta$ (see Figure \ref{fig:Fig2}(c)). 
This is due to the fact that, in this sector with three spins up, $\Delta_{\alpha,\beta}$ in Eqs. (\ref{eq:mr_matrices}) and (\ref{eq:S_lab}) is independent from $\Delta$ and hence $\rhop_{\rm NESS}$ is also invariant with it. This explains why the total heat current and the local spin current do not change. However, the local heat current operator depends on $\Delta$ via Eqs. (\ref{eq:ham_loc}) and (\ref{eq:jH}), and hence $j^{\rm H}$ changes with $\Delta$.

In Figure \ref{fig:Fig2}(b,d) we show the effect of the interaction $\Delta$ on the currents in the sector with two spins up. In this case, the local heat currents also demonstrate a transition between the three different geometries, from parallel (yellow shaded regime), to clockwise direction (green shaded regime), and to counterclockwise direction (red shaded region). However, the total heat currents vary in a non-monotonous way, with a minimum at $\Delta\approx 2.19$. This minimum corresponds to a sharp change in the geometry of the local heat currents and the local spin current which experience an interaction induced inversion of the direction, from clockwise to counterclockwise. 
This current inversion occurs because of the presence of a small avoided crossing near $\Delta\approx 2.19$ which connects two different energy eigenstates at sufficiently low energy such that they have a large enough weight to play a significant role in the properties of the system.   
We will discuss in more detail the role of avoided crossings in the next section. 

\section{Ergotropy}
\label{Sec:ergotropy}
We now show how interactions can significantly affect the possibility of extracting energy from the system via unitary processes $U$. In this section, we focus on the sector with two spins up. The maximum energy that can be extracted via a unitary process is quantified by the \emph{ergotropy} $\mathcal{E}$ \cite{Allahverdyan2004}, which is calculated for the steady state using  
\begin{align}
	\mathcal{E}=\mathrm{Tr}(\rhop_{\rm NESS} \Hop)-\mathrm{Tr}(\rhop_{\rm passive} \Hop),
\end{align}
where $\rhop_{\rm passive}$ is the corresponding passive state \cite{Pusz1978} which is built from the eigenvalues $p_k$ of $\rhop_{\rm NESS}$, and from the system Hamiltonian in its eigenbasis $\Hop_{\rm S}=\sum_{k\ge 1} E_k \Ket{E_k} \Bra{E_k}$ where $E_1 \le E_2 \le \cdots$. More~precisely, 
\begin{align}
	\rhop_{\rm passive} = \sum_k p_k \Ket{E_k} \Bra{E_k}, ~\text{with}~p_{k+1} \le p_k, 
\end{align}
which means that higher energy levels are less populated. Given a state, the passive state is the lowest energy state that can be reached via unitary transformations. Examples of passive states are thermal states, from which it is not possible to extract energy (work) with unitary operations. 

\begin{figure}
	\centering 
        \includegraphics[width=\columnwidth]{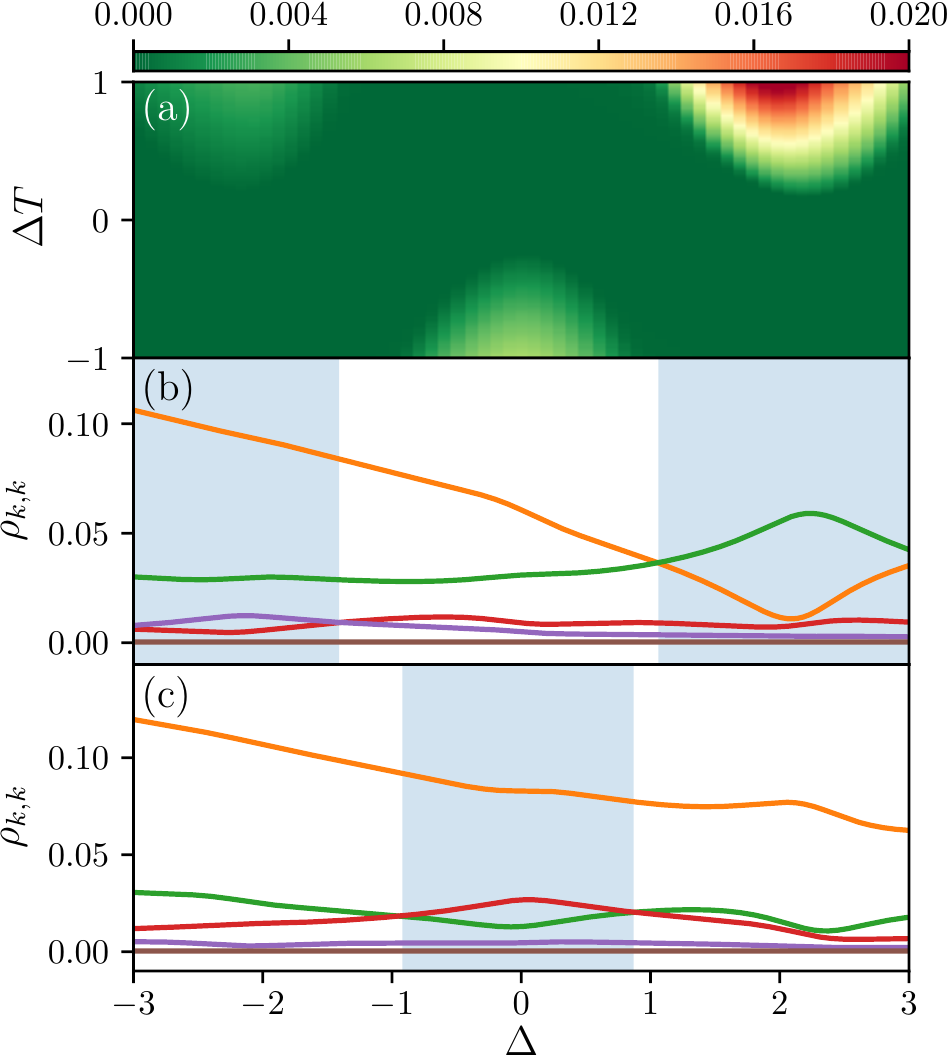}
        \caption{(\textbf{a}) contour plot for ergotropy $\mathcal{E}$ as a function of temperature bias $\Delta T$ and interaction strength $\Delta$ with $(T_{\rm L}+T_{\rm R})/2 = 1.5$; (\textbf{b,c}) occupation of five high energy levels $\rho_{2,2}$ (orange line), $\rho_{3,3}$ (green line), $\rho_{4,4}$ (red line) $\rho_{5,5}$ (purple line), $\rho_{6,6}$ (brown line) as a function of interaction strength $\Delta$ when (\textbf{b}) $T_{\rm L} = 2$ and $T_{\rm R} = 1$ (i.e., $\Delta T=1$) or (\textbf{c}) $T_{\rm L} = 1$ and $T_{\rm R} = 2$ (i.e., $\Delta T=-1$). The shaded regions highlight the portions with population inversion. The system-bath coupling $\gamma=0.01$ for both baths, the local magnetic fields values are $h^z_2=2$ and $h^z_3=1$, and we are considering the sector with two spins up.}
        \label{fig:Fig3}
\end{figure}

In Figure \ref{fig:Fig3}(a), we study the ergotropy extraction within the system due to the temperature bias and the interaction strength. Increasing temperature bias, $\Delta T=T_{\rm L}-T_{\rm R}$, drives the system more out of equilibrium while interaction can be used to tune the energy level differences. For illustration purposes, we consider a fixed average temperature bias $(T_{\rm L}+T_{\rm R})/2 = 1.5$. Various hot spots for ergotropy are identified far from equilibrium as shown in Figure \ref{fig:Fig3}(a). Note that the ergotropy is not symmetric with respect to the interaction $\Delta$ or temperature bias $\Delta T$, indicating a rectification of ergotropy in the~system. 

In general, ergotropy can be non-zero for two main reasons: an occupation of higher energy levels due to a strong temperature bias, or the presence of coherence in the nonequilibrium steady state. From Figure \ref{fig:Fig3}(b,c) we observe that in our setup the main cause is the change in the occupation of the energy levels, which in the following we refer to as ``population inversion''. We note that this is expected within the weak system-bath coupling limit and our perturbative approach which can accurately study systems with small off-diagonal terms in the energy eigenbasis \cite{Fleming2011, Thingna2012, Thingna2013}. 
Figure \ref{fig:Fig3}(b,c) show the occupations of different energy levels $\rho_{k,k}$ (where $k$ increases for increasing energy) versus $\Delta$. When comparing Figure \ref{fig:Fig3}(b) with the line $\Delta T =1$ in Figure \ref{fig:Fig3}(a), or Figure \ref{fig:Fig3}(c) with the line $\Delta T =-1$ in Figure \ref{fig:Fig3}(a), we observe that the ergotropy maxima correspond to the regions with population inversion, highlighted by the light blue shadings in Figure \ref{fig:Fig3}(b,c). In particular, in Figure \ref{fig:Fig3}(b), there is a cross over between $\rho_{2,2}$ (orange line) and $\rho_{3,3}$ (green line) for larger $\Delta$ and between $\rho_{4,4}$ (red line) and $\rho_{5,5}$ (purple line) for lower $\Delta$. In Figure \ref{fig:Fig3}(c), the population inversion is between $\rho_{3,3}$ (green line) and $\rho_{4,4}$ (red line). This confirms that the leading contribution to ergotropy is from population inversion.

\begin{figure}
	\centering 
        \includegraphics[width=\columnwidth]{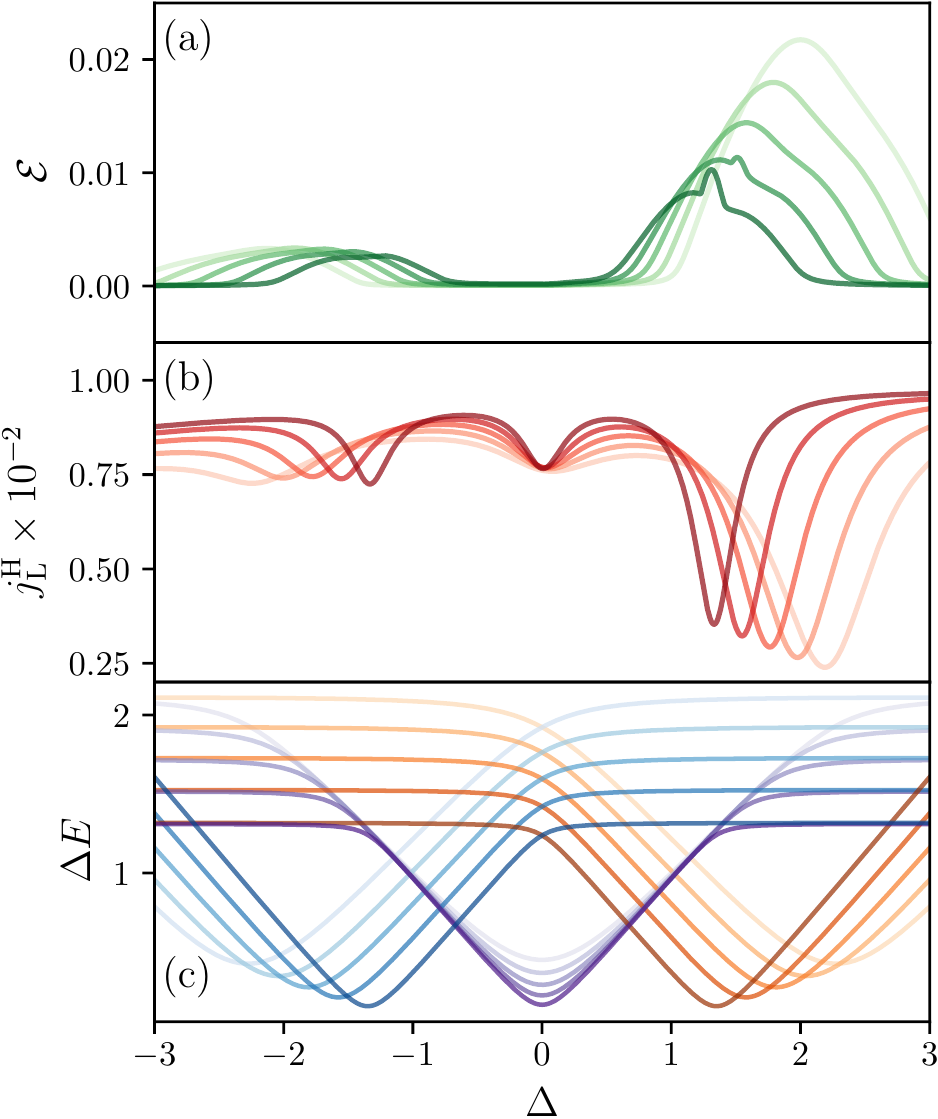}
        \caption{(\textbf{a}) ergotropy (green solid line) versus interaction $\Delta$ for different values of the local magnetic fields $h^z_l$; (\textbf{b}) heat current exchanged with the baths (red solid line) as a function of interaction strength $\Delta$ for different values of the local magnetic fields $h^z_l$; (\textbf{c}) energy difference as a function of interaction $\Delta$ for low energy states $E_3-E_2$ (orange lines), intermediate energy states $E_4-E_3$ (purple lines), and high energy states $E_5-E_4$ (blue lines). In (\textbf{a}--\textbf{c}), color gradient indicates the strength of the local magnetic field. $h^z_2/h^z_3=2$ where $h^z_2=2,~1.8,~1.6,~1.4,~1.2$ for increasingly dark colors. $T_{\rm L} = 2$, and $T_{\rm R} = 1$. The system-bath coupling $\gamma$=0.01 for both baths, and we are considering the sector with two spins up.}
        \label{fig:Fig4}
\end{figure}

In the following, we show that the population inversion occurs \emph{close} to avoided crossings due to the presence of smaller energy gaps \footnote{We emphasize that the avoided crossings and the ergotropy maxima do not exactly coincide due to small corrections from other terms in the master equation}~\eqref{eq:mr_matrices}. This gives a mechanism to use interactions to tune the amount of ergotropy in the system.  
In Figure~\ref{fig:Fig4}(a,b), we show the ergotropy $\mathcal{E}$ and the total heat current $j^{\rm H}$ versus $\Delta$ for different values of the local magnetic fields $h^z_l$. For both $\mathcal{E}$ and $j^{\rm H},$ we observe peaks which shift due to the interplay between the interaction and the local magnetic fields. 
With Figure \ref{fig:Fig4}(c), we can associate these peaks with the avoided crossings. In Figure \ref{fig:Fig4}(c), we show the difference in energy between levels $E_3$ and $E_2$ (orange lines), levels $E_4$ and $E_3$ (purple lines) and levels $E_5$ and $E_4$ (blue lines) versus $\Delta$ and for the same local magnetic fields used in Figure \ref{fig:Fig4}(a,b). It is clear that the maxima of ergotropy and the minima of heat current occur together with the small avoided crossings. Moreover, Figure \ref{fig:Fig3}(b) shows that avoided crossings in low energy states ($\rho_{3,3}$ and $\rho_{4,4}$) result in larger ergotropy as these levels are more occupied than those with higher~energy. 

\section{\label{Sec:conclusions} Conclusions}
We have considered a minimal system to study the interplay between heat and spin currents. In particular, our setup is composed of four spins in a ring which is connected to two baths at different temperatures. Each spin is interacting with its nearest neighbor and a position-dependent magnetic field breaks the reflection and rotation symmetries of the system. {This setup could be implemented with ultracold ions realizations of spin-$1/2$ systems \cite{PorrasCirac2004, BermudezPorras2011, BermudezPlenio2013}, or with quantum dots circuits \cite{RoggeHaug2008, ThalineauMeunier2012, SeoMahalu2013}}. Even without interactions, it is possible for the heat current in the system to flow in three different ways: clockwise, counterclockwise, and in parallel flows. However, we show here that interactions give us a means to cause sizeable changes in the total heat current and induce an inversion in the spin current. When coupling the system with an ancilla, such changes in internal currents could be used for sensing or activation. The internal currents can also be used as indicators of quantum criticality \cite{VoglBrandes2012, SchallerBrandes2014}.   

Coupling such a system to two baths brings it to a nonequilibrium steady state from which it is possible to extract work with unitary processes, i.e., the system can have non-zero ergotropy. We show that the ergotropy is, in general, non-symmetric with respect to the temperature bias $\Delta T$, and it can be significantly enhanced close to avoided crossings due to population inversions. 

For larger rings, the number of relevant avoided crossings can increase; however, we expect the same qualitative behavior. Systems with more complex topologies, e.g., two or more rings, can present a larger variety of behaviors which is worth investigating further.  
In future works, similarly to \cite{Bissbort2017}, we may consider the effect of time-independent and time-dependent gauge fields and consider the thermodynamic properties of such systems. In our setup, the weak coupling between the system and the baths limits the contribution of coherence to the ergotropy. It would thus be interesting to study the effects of stronger system-bath coupling on heat currents and ergotropy.

\section*{ACKNOWLEDGMENTS}
D.P. acknowledges support from the Ministry of Education of Singapore AcRF MOE Tier-II (project MOE2016-T2-1-065, WBS R-144-000-350-112).

\bibliographystyle{apsrev4-1}

\end{document}